\documentclass[aps,prd,twocolumn,groupaddress,tighten,nofootinbib]{revtex4}
\usepackage{amssymb}
\usepackage{mathrsfs}
\usepackage{mathbbold}
\usepackage{graphicx}
\usepackage{amsmath}
\usepackage{epsfig}
\usepackage{color}
\usepackage{mciteplus}
\def\slc#1{\setbox0=\hbox{$#1$}           
    \dimen0=\wd0                                 
    \setbox1=\hbox{/} \dimen1=\wd1               
    \ifdim\dimen0>\dimen1                        
       \rlap{\hbox to \dimen0{\hfil/\hfil}}      
       #1                                        
    \else                                        
       \rlap{\hbox to \dimen1{\hfil$#1$\hfil}}   
       /                                         
    \fi}

\begin{document}

\draft

\title{Non-unitarity of the leptonic mixing matrix in the TeV-scale type-I seesaw model}

\author{Tommy Ohlsson}
\email{tommy@theophys.kth.se}

\author{Christoph Popa}
\email{cgpopa@kth.se}

\author{He Zhang}
\email{zhanghe@kth.se}

\affiliation{Department of Theoretical Physics, School of
Engineering Sciences, Royal Institute of Technology (KTH) --
AlbaNova University Center, Roslagstullsbacken 21, 106 91 Stockholm,
Sweden}

\begin{abstract}
The non-unitarity effects in leptonic flavor mixing are regarded as
one of the generic features of the type-I seesaw model. Therefore,
we explore these effects in the TeV-scale type-I seesaw model, and
show that there exist non-trivial correlations among the
non-unitarity parameters, stemming from the typical flavor structure
of the low-scale seesaw model. In general, it follows from
analytical discussions and numerical results that all the six
non-unitarity parameters are related to three model parameters,
while the widely studied parameters $\eta_{e\tau}$ and
$\eta_{\mu\tau}$ cannot be phenomenologically significant
simultaneously.
\end{abstract}
\maketitle

\section{Introduction}
\label{sec:intro}

During the past decade, experimental progress on neutrino masses and
leptonic mixing has opened up a new window in searching for physics
beyond the Standard Model (SM) of particle physics. Since neutrinos
are massless particles in the SM, one usually extends the SM
particle content in order to accommodate massive neutrinos. Among
various theories of this kind, the seesaw
mechanism~\cite{Minkowski:1977sc,*Yanagida:1979as,*GellMann:1980vs,*Mohapatra:1979ia}
attracts a lot of attention in virtue of its naturalness and
simplicity. In the conventional type-I seesaw model, three
right-handed neutrinos are introduced and assigned large Majorana
masses. In order to stabilize the masses of the light neutrinos at
the sub-eV scale, the masses of right-handed neutrinos are usually
chosen to be close to the Grand Unified (GUT) scale, i.e., $10^{16}~
{\rm GeV}$ far above the electroweak scale $\Lambda_{\rm EW} \sim
100 ~{\rm GeV}$. Thus, the typical type-I seesaw model suffers from
lack of testability, since right-handed neutrinos are too heavy to
be produced in current collider experiments. The Large Hadron
Collider (LHC) will soon bring us a revolution in particle physics
at the TeV-scale, which leads neutrino physics at the TeV-scale to
an exciting direction. In this respect, the question, if we can find
the answer of neutrino mass generation at the LHC, draws more and
more attention.

In order for the type-I seesaw model to be testable, i.e., bringing
the right-handed neutrino masses down to the TeV level, the small
neutrino masses have to be effectively suppressed via other
mechanisms rather than the GUT scale, such as radiative generation,
small lepton number breaking, or neutrino masses from a higher than
dimension-five effective operator (see,
e.g.~Ref.~\cite{Bonnet:2009ej} and references therein). Besides,
special cancellations in the contributions to the neutrino masses
can also be employed to solve this problem. For instance, in the
type-I+II seesaw model, one may assume the right-handed neutrino
contributions to the light neutrino mass matrix to be comparable
with the contributions originated from the triplet Higgs field, and
the tiny left-handed neutrino masses are suppressed if there is a
severe cancellation between these two mass terms. However, such a
scheme seems implausible, since it involves strong fine-tuning
between different and unrelated sources.

In this work, we will focus on a simple, but realistic, low-scale
type-I seesaw model, in which a structural cancellation among the
contributions from different right-handed neutrinos plays the key
role of protecting neutrino
masses~\cite{Pilaftsis:1991ug,*Kersten:2007vk,*Zhang:2009ac,*Adhikari:2010yt}.
Such a structural cancellation could naturally be regarded as the
appearance of certain flavor symmetries stemming from some
underlying dynamics. Furthermore, the mixing between heavy and light
neutrinos results in observable non-unitarity (NU) effects in
leptonic flavor
mixing~\cite{Antusch:2006vwa,*Antusch:2008tz,Tang:2009na,*Meloni:2009cg,FernandezMartinez:2007ms,*Xing:2007zj,*Goswami:2008mi,*Xing:2008fg,*Luo:2008vp,*Altarelli:2008yr,*Malinsky:2009gw,*Blennow:2009pk,*Antusch:2009pm,*Rodejohann:2009cq,*Malinsky:2009df,*He:2009ua,*Xing:2009ce,*Chao:2007mz,*Dev:2009aw,*Antusch:2009gn,*Rodejohann:2009ve},
which are usually parametrized by using so-called NU parameters. In
what follows, we will discuss the possible NU effects in this
framework in detail, and in particular, we will show that, due to
the special flavor structure of the TeV-scale type-I seesaw,
non-trivial correlations among the NU parameters exist, and not all
the NU parameters can be phenomenologically significant in future
neutrino oscillation experiments.

The remaining part of this work is organized as follows. In
Sec.~\ref{sec:model}, we introduce the low-scale type-I seesaw
model. Then, in Sec.~\ref{sec:NU}, we explore in detail the typical
features of the NU parameters in the model. A discussion on the
possible effects in future neutrino oscillation experiments is also
given. Section~\ref{sec:nums} is devoted to numerical illustrations
of the allowed parameter regions. Finally, a brief summary is
presented in Sec.~\ref{sec:summary}.

\section{low-scale type-I seesaw model}
\label{sec:model}

We first write out explicitly the Lagrangian responsible for
neutrino Yukawa interactions and the Majorana mass term of
right-handed neutrinos in the type-I seesaw model, viz.,
\begin{eqnarray}\label{L}
{\cal L} = -\overline{\ell_{L\rm}} \tilde\phi Y^\dagger_\nu \nu_{\rm
R} - \frac{1}{2} \overline{\nu_{\rm R}} M_{\rm R} \nu^c_{\rm R} +
{\rm h.c.} \, ,
\end{eqnarray}
with $\tilde \phi = {\rm i} \tau_2 \phi^*$, where $\ell_{\rm L}$,
$\nu_{\rm R}$, and $\phi$ denote lepton doublets, right-handed
neutrinos, and Higgs fields, respectively. Here $Y_\nu$ and $M_{\rm
R}$ stand for the corresponding Yukawa couplings and right-handed
neutrino mass matrix. If the right-handed neutrino scale is higher
than the electroweak scale, one should integrate out the heavy
degree of freedom of right-handed components in dealing with
low-scale processes. Explicitly, at tree level, the only
dimension-five operator is the Weinberg operator,
\begin{eqnarray}\label{L5}
{\cal L}_W = \left( Y^T_\nu \frac{1}{M_{\rm R}}
Y_\nu\right)_{\alpha\beta} \left( \overline{\ell_{{\rm L}\alpha}}
\varepsilon \tau^i \ell^c_{{\rm L}\beta} \right)\left( \tilde\phi^T
\varepsilon \tau^i \tilde\phi \right) + {\rm h.c.}
\end{eqnarray}
After electroweak symmetry breaking, the SM Higgs field acquires a
nonzero vacuum expectation value $v\simeq174~{\rm GeV}$, while the
Majorana mass matrix of the light neutrinos then reads
\begin{eqnarray}\label{mnu}
m_\nu = -{v^2} Y^T_\nu \frac{1}{M_{\rm R}} Y_\nu \, .
\end{eqnarray}
By defining $M_{\rm D} = vY_\nu$, Eq.~\eqref{mnu} reproduces the
ordinary type-I seesaw formula.

In the normal type-I seesaw model, the masses of the light neutrinos
are purely suppressed by the ratio of the electroweak scale and the
$B-L$ breaking scale, i.e., $v/M_{\rm R}$, and hence, the
right-handed neutrino masses are usually chosen to be close to the
GUT scale. In order to lower the $B-L$ scale to the scope of current
colliders, i.e., the TeV-scale, one can simply assume the neutrino
Yukawa couplings to be much smaller than those of other SM fermions.
However, this will result in tiny couplings between the right-handed
neutrinos and the charged leptons. Consequently, the production
cross section of heavy neutrinos is also negligible. In this
respect, one would like to maintain sizable Yukawa couplings and
keep the neutrino masses stable simultaneously. In order to achieve
this goal in the type-I seesaw framework, the contributions to
neutrino masses from different right-handed neutrinos have to
cancel, at least at leading order. The small neutrino masses can
then be viewed as perturbations to the structure cancellation.

The necessary and sufficient conditions for such an exact
cancellation require the following form of the Yukawa
couplings~\cite{Pilaftsis:1991ug,*Kersten:2007vk,*Zhang:2009ac,*Adhikari:2010yt}
\begin{eqnarray}\label{Y}
Y_\nu = \left(\begin{matrix} x_1 & ax_1 & b x_1 \cr x_2 & a x_2 & b
x_2 \cr x_3 & a x_3 & b x_3
\end{matrix}\right) \ ,
\end{eqnarray}
where $a$ and $b$ are free parameters, and the relation between
right-handed neutrino masses is given by
\begin{eqnarray}\label{MN}
\frac{x^2_1}{M_1} + \frac{x^2_2}{M_2} + \frac{x^2_3}{M_3} = 0 \ ,
\end{eqnarray}
where we have already chosen a basis in which the right-handed
neutrino mass matrix is diagonal. Using Eq.~\eqref{mnu}, one can
easily prove that, under the above conditions, the neutrino masses
vanish, namely, $m_\nu=0$. Note that the radiative corrections
induced by right-handed neutrinos, such as renormalization group
running effects, may spoil the stability of small neutrino masses
unless the right-handed neutrinos are nearly degenerate in mass as
required by some flavor symmetric theories.

\section{Non-unitarity of the leptonic mixing matrix}
\label{sec:NU}

Besides the dimension-five operator discussed above, there exists a
unique dimension-six
operator~\cite{Broncano:2002rw,*Broncano:2003fq,*Abada:2007ux,*Gavela:2009cd}
\begin{eqnarray}\label{eq:L5}
{\cal L}_6 = C^6_{\alpha\beta}  \left( \overline{\ell_{{\rm
L}\alpha}} \tilde\phi \right) {\rm i} \slc{\partial}
\left(\tilde\phi^\dagger \ell_{{\rm L}\beta} \right) \, ,
\end{eqnarray}
where the coefficient is given by
\begin{eqnarray}\label{eq:C6}
C^6 =Y^\dagger_\nu \frac{1}{M^\dagger_{\rm R}} \frac{1}{M_{\rm R}}
Y_\nu \, .
\end{eqnarray}
After the electroweak symmetry breaking, the dimension-six operator
given in Eq.~\eqref{eq:L5} leads to corrections to the kinetic
energy terms for the light neutrinos. Therefore, in order to keep
the neutrino kinetic energy canonically normalized, one has to
rescale the neutrino fields by using the following transformation
\begin{eqnarray}\label{scale}
\nu'_{{\rm L}\alpha} = \left(\delta_{\alpha\beta} + v^2
C^6_{\alpha\beta} \right)^{\frac{1}{2}} \nu_{{\rm L}\beta} \ .
\end{eqnarray}
Due to this field rescaling, the usual leptonic mixing matrix $U$,
which relates the neutrino flavor basis and mass basis, is replaced
by a non-unitary matrix as
\begin{eqnarray}\label{N}
N=\left(1 - \frac{ v^2}{2} C^6 \right) U = \left(1 + \eta \right) U
= RU \, ,
\end{eqnarray}
where $\eta$ is a Hermitian matrix containing totally nine
parameters, i.e., six moduli and three phases governing the NU
effects, and $U$ diagonalizes the light neutrino mass matrix as $
U^\dagger m_\nu U^* ={\rm diag}(m_1,m_2,m_3)$ with $m_i$ being the
masses of the light neutrinos.

Note that, different from the dimension-five operator, under the
assumptions in the previous section, the dimension-six operator is
not necessarily vanishing, since the flavor structure is different
and $C^6$ is suppressed by the square of $M_{\rm R}$. Combining
Eqs.~(\ref{Y})-(\ref{MN}) and (\ref{N}), we can explicitly write
down the NU parameters as
\begin{eqnarray}\label{eta}
\eta = -\frac{v^2}{2} C^6= \eta_0 \left(\begin{matrix} 1 & a & b \cr
a^* & |a|^2 & a^* b \cr b^* & ab^* & |b|^2
\end{matrix}\right) \ ,
\end{eqnarray}
where
\begin{eqnarray}\label{eta0}
\eta_0  =  -\frac{v^2}{2}\left( \frac{|x_1|^2}{M^2_1} +
\frac{|x_2|^2}{M^2_2} + \frac{|x_3|^2}{M^2_3} \right) \, .
\end{eqnarray}
As a rough estimate, if we choose $M_i \sim {\rm TeV}$ and the
Yukawa couplings at order one, then $\eta \sim 0.1~\%$ can be
expected. In addition, the magnitudes of the NU parameters are
constrained from universality test, rare lepton decays, and
invisible width of $Z$-boson. The present bounds at 90~\% C.L. on
the NU parameters are given
by~\cite{Antusch:2006vwa,*Antusch:2008tz}
\begin{eqnarray}\label{eq:rhoB}
|\eta| < \left(\begin{matrix} 2.0\times10^{-3} & 6.0 \times 10^{-5}
& 1.6 \times10^{-3} \cr \sim & 8.0\times10^{-4} & 1.1 \times10^{-3}
\cr \sim & \sim & 2.7\times10^{-3}
\end{matrix}\right)\ ,
\end{eqnarray}
in which, the most severe constraint is that on the $e\mu$ element
coming from the $\mu\to e\gamma$ decay.\footnote{Note that, in
deriving these constraints, the condition $M_{\rm R}>\Lambda_{\rm
EW}$ is assumed. In case of $M_{\rm R} < \Lambda_{\rm EW}$, the
constraint on $\eta_{e\mu}$ is relaxed due to the restoration of the
GIM mechanism.}

In the literature, these NU parameters are usually taken as free
parameters. However, according to Eq.~\eqref{eta}, the NU parameters
are not independent in general. The correlations between these
parameters stemming from some possible flavor symmetries can be
viewed as a typical feature of the low-scale type-I seesaw model,
and should be tested in the future neutrino oscillation experiments.

Concretely, it can be seen from Eq.~\eqref{eta} that the NU
parameters are governed by the three independent parameters
$\eta_0$, $a$, and $b$. The present restrictions on the elements of
$\eta$ are at percentage level, except a rather stringent bound
$\eta_{e\mu} < 6.0 \times 10^{-5}$. Hence, to avoid severe unitarity
constraints, one may expect either $a$ or $\eta_0$ in
Eq.~\eqref{eta} to be tiny. If $a$ is very small, one may ignore the
NU parameters proportional to $a$, and Eq.~\eqref{eta} can be
simplified to
\begin{eqnarray}\label{etaS}
\eta \simeq \eta_0 \left(\begin{matrix} 1 & 0 & b \cr 0 & 0 & 0 \cr
b^* & 0 & |b|^2
\end{matrix}\right) \ .
\end{eqnarray}
In this limit, both the $\eta_{e\mu}$ and $\eta_{\mu\tau}$ cannot be
significant and $\eta_{e\tau}$ is the only possibly large NU
parameter. On the other hand, in the case $\eta_0$ is very small
while $a$ and $b$ are relatively large, the first row and column in
$\eta$ can be ignored, and one has approximately
\begin{eqnarray}\label{etaS2}
\eta \simeq \eta_0 \left(\begin{matrix} 0 & 0 & 0 \cr 0 & |a|^2 &
a^*b \cr 0 & ab^* & |b|^2
\end{matrix}\right) \ .
\end{eqnarray}
Therefore, $\eta_{e\tau}$ and $\eta_{\mu\tau}$ cannot be sizable
simultaneously, although their current upper bounds are both within
the sensitivity scope of a neutrino factory.

When neutrinos propagate in vacuum, in the ultra-relativistic limit
$E\gg m_{i}$, the time evolution in the flavor basis is described by
the Hamiltonian
\begin{eqnarray}\label{H}
H  =  \frac{1}{2E} \tilde R^* \left[ U^* \cdot  {\rm diag}
\left(m^2_1,m^2_2, m^2_3\right) \cdot  U^T \right] (\tilde R^*)^{-1}
\, ,
\end{eqnarray}
where the normalized $\tilde{R}_{\alpha\beta} \equiv R_{\alpha
\beta}\left(RR^\dagger\right)^{-\frac{1}{2}}_{\alpha\alpha}$ is used
instead of $R_{\alpha\beta}$ for the consistency between quantum
states and fields. The transition amplitude from a neutrino flavor
$\alpha$ to another neutrino flavor $\beta$ after traveling a
distance $L$ can now be obtained
as~\cite{Ohlsson:2008gx,*Meloni:2009ia}
\begin{eqnarray}\label{A}
A_{\alpha\beta} (L)  =    \sum_i F^i_{\alpha\beta} \exp\left({-{\rm
i} \frac{m^2_i L}{2E}}\right) \, ,
\end{eqnarray}
where $F^i_{\alpha\beta} = \sum_{\gamma ,\rho} (\tilde R^*)_{\alpha
\gamma } (\tilde R)_{\beta \rho}U^*_{\gamma i} U_{\rho i}$. With the
above definitions, the oscillation probability is then given by $
P_{\alpha\beta} (L) \equiv \left| {A}_{\alpha\beta}(L) \right|^2 $.
A salient feature is that, in the case $\alpha \neq \beta$,
$P_{\alpha\beta}(0)$ is not vanishing generally. Therefore, a flavor
transition might already happen at the source even before the
oscillation process, which is known as the zero-distance effect.

In general, a near detector at a short distance provides the best
sensitivities to the NU parameters, since the standard oscillation
effects in the unitary limit are suppressed with respect to the
baseline length. In particular, in a future neutrino factory, a near
detector with $\nu_\tau$ detection is shown to be useful for
studying NU effects \cite{Tang:2009na}. In this respect, one may be
interested in the flavor transitions in the appearance channels. For
short enough distances, the oscillation amplitudes approximate to
\begin{eqnarray}\label{Aapp}
A_{\alpha\beta}(L) \simeq  A^{\rm SM}_{\alpha\beta}(L) +
2\eta^*_{\alpha\beta} \, ,
\end{eqnarray}
where $A^{\rm SM}_{\alpha\beta}(L)$ denotes the oscillation
amplitude of the unitary analysis. With respect to the NU parameters
in the present model, i.e., Eqs.~\eqref{etaS} and \eqref{etaS2}, the
channels $\nu_e \to \nu_\tau$ and $\nu_\mu \to
\nu_\tau$~\cite{Donini:2002rm,*Donini:2008wz} turn out to be the
best options to search for the NU effects. On the other hand, if
sizable NU effects are observed in both two channels, the simplest
type-I seesaw model will be ruled out. The possible way out might
then be theories possessing more than three heavy neutrinos, e.g.
the inverse seesaw model~\cite{Mohapatra:1986bd} or theories with
extra spatial dimensions~\cite{Bhattacharya:2009nu,*Blennow:2010zu}.

\section{numerical analysis}\label{sec:nums}

We proceed to perform a full scan of the parameter space of the
model in order to obtain predictions for the NU parameters. For each
set of these parameters, we compare the model predictions to the
experimental data with a $\chi^2$ function
\begin{eqnarray}\label{eq:chi2}
\chi^2 = \sum_i \frac{(\rho_i - \rho^0_i)^2}{\sigma^2_{i}} \ ,
\end{eqnarray}
where $\rho^0_i$ are assumed to be zero for the central values of
the NU parameters, $\sigma_i$ the corresponding 1$\sigma$ absolute
error, and $\rho_i$ the predicted values of $\eta$'s. In our
numerical analysis, we make use of the current bounds given in
Eq.~\eqref{eq:rhoB}. Note that, since neutrino masses are assumed to
be generated via other mechanisms (e.g.~deviations from the exact
structure cancellations), we do not consider experimental
constraints on neutrino masses and leptonic flavor mixing parameters
in our analysis. Discussions on the neutrino mass generation in the
current framework can be found in
Refs.~\cite{Pilaftsis:1991ug,*Kersten:2007vk,*Zhang:2009ac,*Adhikari:2010yt}.

\begin{figure}[t]
\begin{center}\vspace{-0.2cm}
\includegraphics[width=5.5cm,bb=150 200 750 800]{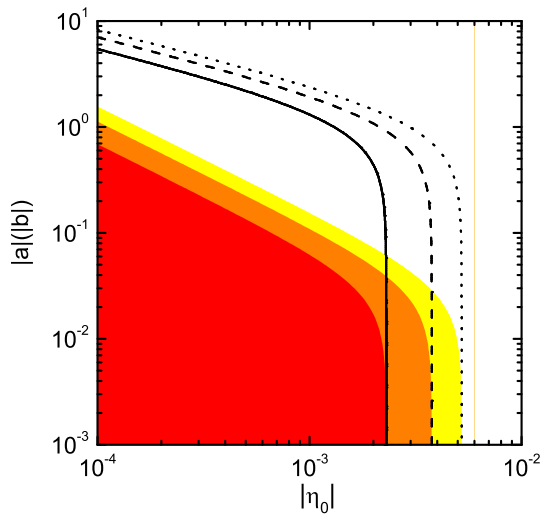}
\vspace{1.5cm} \caption{\label{fig:fig1} Allowed regions of the
model parameters $a$ (colored regions), $b$ (curves), and $\eta_0$
at 1$\sigma$ (red, solid), 2$\sigma$ (orange, dashed), and 3$\sigma$
(yellow, dotted) C.L.}\vspace{-0.6cm}
\end{center}
\end{figure}

In Fig.~\ref{fig:fig1}, we show the allowed parameter space of $a$,
$b$, and $\eta_0$ at 1$\sigma$, 2$\sigma$, and 3$\sigma$ C.L.
One can observe from the plot that there exists an upper bound on
$\eta_0$, which results from the relation $\eta_0 = \eta_{ee}$
according to Eq.~\eqref{eta}. Furthermore, for a fixed $\eta_0$, the
allowed regions of $b$ are larger than those of $a$, i.e., the upper
bound of $b$ is about one order of magnitude larger than that of
$a$. This is in agreement with our analytical analysis, since
$\eta_{e\mu}$ sets only a strong constraint on $a$ but not on $b$.
For a very tiny $\eta_0$, $a$ and $b$ can be arbitrarily chosen
without suffering from stringent unitarity constraints, i.e., their
upper bounds approach infinity. However, as we mentioned before, for
a realistic type-I seesaw model with sub-eV scale neutrino masses,
these parameters cannot be arbitrarily small unless another
mechanism responsible for the masses of the light neutrinos is
considered.

\begin{figure}[t]
\begin{center}\vspace{-1.5cm}
\includegraphics[width=11.3cm,bb=0 -150 600 450]{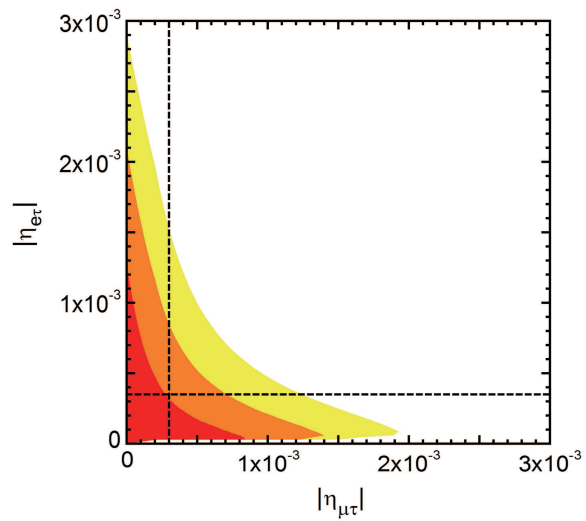}
\vspace{-3.4cm} \caption{\label{fig:fig2} Allowed regions of the NU
parameters $|\eta_{e\tau}|$ and $|\eta_{\mu\tau}|$ at 1$\sigma$
(red), 2$\sigma$ (orange), and 3$\sigma$ (yellow) C.L. The dashed
lines in the figure correspond to the sensitivities to these NU
parameters (at 90~\% C.L.) in a near OPERA-like tau-detector of a
neutrino factory (parent muon energy $E=25~{\rm GeV}$).
}\vspace{-0.5cm}
\end{center}
\end{figure}

According to Eq.~\eqref{Aapp}, the phenomenologically interesting NU
parameters are the off-diagonal elements in $\eta$. Since
$\eta_{e\mu}$ is strongly constrained experimentally, the remaining
ones are $\eta_{e\tau}$ and $\eta_{\mu\tau}$, whose allowed
regions\footnote{Given the fact that there is no experimental
information on the leptonic CP-violation until now, we do not
include the CP-violating phases of $\eta$'s, and only show the
constraints on the absolute values of these NU parameters.} are
illustrated in Fig.~\ref{fig:fig2}.
One can observe that both $\eta_{e\tau}$ and $\eta_{\mu\tau}$ can
reach their upper bounds in general. However, as we expected from
the above analysis, in order for one of them to be sizable, the
other one has to be suppressed, reflecting the underlying
correlations among the NU parameters. For example, in the case of
$\eta_{e\tau} > 10^{-3}$, a severe bound $\eta_{\mu\tau} \lesssim 3
\times 10^{-4}$ (at 2$\sigma$ C.L.) can be read off from the plot.
Similarly, in the case of $\eta_{\mu\tau} > 10^{-3}$, one has
$\eta_{e\tau} \lesssim 3 \times 10^{-4}$ (at 2$\sigma$ C.L.). For
comparison, in the plot we also show the discovery potential of the
NU parameters by using an OPERA-like near tau-detector of a future
neutrino factory (see detailed discussions on the detector setup in
e.g.~Refs.~\cite{Tang:2009na,*Meloni:2009cg}). The regions on the
right-hand side of the vertical line and above the horizontal line
can be well searched for with such an experimental setup. Therefore,
besides the search of the NU effects, it can also shed some light on
distinguishing the possible new physics behind the NU effects.

\section{summary and conclusion}\label{sec:summary}

In this Letter, we have studied the NU effects from the low-scale
type-I seesaw model. We have pointed out that in the realistic
low-scale type-I seesaw model without unacceptable lepton number
violations, there exists a non-trivial flavor structure of the NU
parameters, which originates from the structural cancellations of
the neutrino Yukawa couplings. The underlying correlations among the
NU parameters have been established, and the allowed parameter
spaces have been illustrated. In view of the current constraints on
the NU parameters, we have found that, in the low-scale type-I
seesaw model, there exists only one phenomenologically interesting
NU parameter, i.e., either $\eta_{e\tau}$ or $\eta_{\mu\tau}$. This
flavor structure can be viewed as a distinctive feature of the
low-scale type-I seesaw model, and can be tested in the future
experiments, in particular, in the near tau-detector of a neutrino
factory. In addition to the neutrino oscillation experiments, direct
searches of right-handed neutrinos at colliders also rely on the NU
parameters. Especially, the signals of tri-lepton final states with
transverse missing energy at the LHC provide us with the capability
of revealing the underlying nature of right-handed neutrinos. In
conclusion, the low-scale type-I seesaw features a very distinctive
flavor structure, and a combined analysis of both collider
signatures and neutrino oscillation experiments will be very useful
to obtain knowledge on the physics behind the right-handed
neutrinos.

\begin{acknowledgments}

This work was supported by the Swedish Research Council
(Vetenskapsr{\aa}det), contract no. 621-2008-4210 (T.O.) and the
Royal Institute of Technology (KTH), project no. SII-56510 (H.Z.).
H. Zhang acknowledges Xiangdong Ji for the hospitality at Peking
University and Shanghai Jiaotong University where part of this work
was performed.

\end{acknowledgments}

\bibliography{bib}

\ifx\mcitethebibliography\mciteundefinedmacro
\PackageError{apsrevM.bst}{mciteplus.sty has not been loaded}
{This bibstyle requires the use of the mciteplus package.}\fi
\begin{mcitethebibliography}{41}
\expandafter\ifx\csname natexlab\endcsname\relax\def\natexlab#1{#1}\fi
\expandafter\ifx\csname bibnamefont\endcsname\relax
  \def\bibnamefont#1{#1}\fi
\expandafter\ifx\csname bibfnamefont\endcsname\relax
  \def\bibfnamefont#1{#1}\fi
\expandafter\ifx\csname citenamefont\endcsname\relax
  \def\citenamefont#1{#1}\fi
\expandafter\ifx\csname url\endcsname\relax
  \def\url#1{\texttt{#1}}\fi
\expandafter\ifx\csname urlprefix\endcsname\relax\def\urlprefix{URL }\fi
\providecommand{\bibinfo}[2]{#2}
\providecommand{\eprint}[2][]{\url{#2}}

\bibitem[{\citenamefont{Minkowski}(1977)}]{Minkowski:1977sc}
\bibinfo{author}{\bibfnamefont{P.}~\bibnamefont{Minkowski}},
  \bibinfo{journal}{Phys. Lett.} \textbf{\bibinfo{volume}{B67}},
  \bibinfo{pages}{421} (\bibinfo{year}{1977})\relax
\mciteBstWouldAddEndPuncttrue
\mciteSetBstMidEndSepPunct{\mcitedefaultmidpunct}
{\mcitedefaultendpunct}{\mcitedefaultseppunct}\relax
\EndOfBibitem
\bibitem[{\citenamefont{Yanagida}(1979)}]{Yanagida:1979as}
\bibinfo{author}{\bibfnamefont{T.}~\bibnamefont{Yanagida}}, in
  \emph{\bibinfo{booktitle}{Proc. Workshop on the baryon number of the Universe
  and unified theories}}, edited by
  \bibinfo{editor}{\bibfnamefont{O.}~\bibnamefont{Sawada}} \bibnamefont{and}
  \bibinfo{editor}{\bibfnamefont{A.}~\bibnamefont{Sugamoto}}
  (\bibinfo{year}{1979}), p.~\bibinfo{pages}{95}\relax
\mciteBstWouldAddEndPuncttrue
\mciteSetBstMidEndSepPunct{\mcitedefaultmidpunct}
{\mcitedefaultendpunct}{\mcitedefaultseppunct}\relax
\EndOfBibitem
\bibitem[{\citenamefont{Gell-Mann et~al.}(1979)\citenamefont{Gell-Mann, Ramond,
  and Slansky}}]{GellMann:1980vs}
\bibinfo{author}{\bibfnamefont{M.}~\bibnamefont{Gell-Mann}},
  \bibinfo{author}{\bibfnamefont{P.}~\bibnamefont{Ramond}}, \bibnamefont{and}
  \bibinfo{author}{\bibfnamefont{R.}~\bibnamefont{Slansky}}, in
  \emph{\bibinfo{booktitle}{Supergravity}}, edited by
  \bibinfo{editor}{\bibfnamefont{P.}~\bibnamefont{van Nieuwenhuizen}}
  \bibnamefont{and} \bibinfo{editor}{\bibfnamefont{D.}~\bibnamefont{Freedman}}
  (\bibinfo{year}{1979}), p. \bibinfo{pages}{315}\relax
\mciteBstWouldAddEndPuncttrue
\mciteSetBstMidEndSepPunct{\mcitedefaultmidpunct}
{\mcitedefaultendpunct}{\mcitedefaultseppunct}\relax
\EndOfBibitem
\bibitem[{\citenamefont{Mohapatra and Senjanovi{\'c}}(1980)}]{Mohapatra:1979ia}
\bibinfo{author}{\bibfnamefont{R.~N.} \bibnamefont{Mohapatra}}
  \bibnamefont{and}
  \bibinfo{author}{\bibfnamefont{G.}~\bibnamefont{Senjanovi{\'c}}},
  \bibinfo{journal}{Phys. Rev. Lett.} \textbf{\bibinfo{volume}{44}},
  \bibinfo{pages}{912} (\bibinfo{year}{1980})\relax
\mciteBstWouldAddEndPuncttrue
\mciteSetBstMidEndSepPunct{\mcitedefaultmidpunct}
{\mcitedefaultendpunct}{\mcitedefaultseppunct}\relax
\EndOfBibitem
\bibitem[{\citenamefont{Bonnet et~al.}(2009)\citenamefont{Bonnet,
  Hern{\'a}ndez, Ota, and Winter}}]{Bonnet:2009ej}
\bibinfo{author}{\bibfnamefont{F.}~\bibnamefont{Bonnet}},
  \bibinfo{author}{\bibfnamefont{D.}~\bibnamefont{Hern{\'a}ndez}},
  \bibinfo{author}{\bibfnamefont{T.}~\bibnamefont{Ota}}, \bibnamefont{and}
  \bibinfo{author}{\bibfnamefont{W.}~\bibnamefont{Winter}},
  \bibinfo{journal}{JHEP} \textbf{\bibinfo{volume}{10}}, \bibinfo{pages}{076}
  (\bibinfo{year}{2009}), \eprint{arXiv:0907.3143}\relax
\mciteBstWouldAddEndPuncttrue
\mciteSetBstMidEndSepPunct{\mcitedefaultmidpunct}
{\mcitedefaultendpunct}{\mcitedefaultseppunct}\relax
\EndOfBibitem
\bibitem[{\citenamefont{Pilaftsis}(1992)}]{Pilaftsis:1991ug}
\bibinfo{author}{\bibfnamefont{A.}~\bibnamefont{Pilaftsis}},
  \bibinfo{journal}{Z. Phys.} \textbf{\bibinfo{volume}{C55}},
  \bibinfo{pages}{275} (\bibinfo{year}{1992}), \eprint{hep-ph/9901206}\relax
\mciteBstWouldAddEndPuncttrue
\mciteSetBstMidEndSepPunct{\mcitedefaultmidpunct}
{\mcitedefaultendpunct}{\mcitedefaultseppunct}\relax
\EndOfBibitem
\bibitem[{\citenamefont{Kersten and Smirnov}(2007)}]{Kersten:2007vk}
\bibinfo{author}{\bibfnamefont{J.}~\bibnamefont{Kersten}} \bibnamefont{and}
  \bibinfo{author}{\bibfnamefont{A.~Y.} \bibnamefont{Smirnov}},
  \bibinfo{journal}{Phys. Rev.} \textbf{\bibinfo{volume}{D76}},
  \bibinfo{pages}{073005} (\bibinfo{year}{2007}),
  \eprint{arXiv:0705.3221}\relax
\mciteBstWouldAddEndPuncttrue
\mciteSetBstMidEndSepPunct{\mcitedefaultmidpunct}
{\mcitedefaultendpunct}{\mcitedefaultseppunct}\relax
\EndOfBibitem
\bibitem[{\citenamefont{Zhang and Zhou}(2010)}]{Zhang:2009ac}
\bibinfo{author}{\bibfnamefont{H.}~\bibnamefont{Zhang}} \bibnamefont{and}
  \bibinfo{author}{\bibfnamefont{S.}~\bibnamefont{Zhou}},
  \bibinfo{journal}{Phys. Lett.} \textbf{\bibinfo{volume}{B685}},
  \bibinfo{pages}{297} (\bibinfo{year}{2010}), \eprint{arXiv:0912.2661}\relax
\mciteBstWouldAddEndPuncttrue
\mciteSetBstMidEndSepPunct{\mcitedefaultmidpunct}
{\mcitedefaultendpunct}{\mcitedefaultseppunct}\relax
\EndOfBibitem
\bibitem[{\citenamefont{Adhikari and Raychaudhuri}(2010)}]{Adhikari:2010yt}
\bibinfo{author}{\bibfnamefont{R.}~\bibnamefont{Adhikari}} \bibnamefont{and}
  \bibinfo{author}{\bibfnamefont{A.}~\bibnamefont{Raychaudhuri}}
  (\bibinfo{year}{2010}), \eprint{arXiv:1004.5111}\relax
\mciteBstWouldAddEndPuncttrue
\mciteSetBstMidEndSepPunct{\mcitedefaultmidpunct}
{\mcitedefaultendpunct}{\mcitedefaultseppunct}\relax
\EndOfBibitem
\bibitem[{\citenamefont{Antusch et~al.}(2006)\citenamefont{Antusch, Biggio,
  Fern{\'a}ndez-Mart{\'i}nez, Gavela, and
  L{\'o}pez-Pav{\'o}n}}]{Antusch:2006vwa}
\bibinfo{author}{\bibfnamefont{S.}~\bibnamefont{Antusch}},
  \bibinfo{author}{\bibfnamefont{C.}~\bibnamefont{Biggio}},
  \bibinfo{author}{\bibfnamefont{E.}~\bibnamefont{Fern{\'a}ndez-Mart{\'i}nez}},
  \bibinfo{author}{\bibfnamefont{M.~B.} \bibnamefont{Gavela}},
  \bibnamefont{and}
  \bibinfo{author}{\bibfnamefont{J.}~\bibnamefont{L{\'o}pez-Pav{\'o}n}},
  \bibinfo{journal}{JHEP} \textbf{\bibinfo{volume}{10}}, \bibinfo{pages}{084}
  (\bibinfo{year}{2006}), \eprint{hep-ph/0607020}\relax
\mciteBstWouldAddEndPuncttrue
\mciteSetBstMidEndSepPunct{\mcitedefaultmidpunct}
{\mcitedefaultendpunct}{\mcitedefaultseppunct}\relax
\EndOfBibitem
\bibitem[{\citenamefont{Antusch
  et~al.}(2009{\natexlab{a}})\citenamefont{Antusch, Baumann, and
  Fern{\'a}ndez-Mart{\'i}nez}}]{Antusch:2008tz}
\bibinfo{author}{\bibfnamefont{S.}~\bibnamefont{Antusch}},
  \bibinfo{author}{\bibfnamefont{J.~P.} \bibnamefont{Baumann}},
  \bibnamefont{and}
  \bibinfo{author}{\bibfnamefont{E.}~\bibnamefont{Fern{\'a}ndez-Mart{\'i}nez}},
  \bibinfo{journal}{Nucl. Phys.} \textbf{\bibinfo{volume}{B810}},
  \bibinfo{pages}{369} (\bibinfo{year}{2009}{\natexlab{a}}),
  \eprint{arXiv:0807.1003}\relax
\mciteBstWouldAddEndPuncttrue
\mciteSetBstMidEndSepPunct{\mcitedefaultmidpunct}
{\mcitedefaultendpunct}{\mcitedefaultseppunct}\relax
\EndOfBibitem
\bibitem[{\citenamefont{Tang and Winter}(2009)}]{Tang:2009na}
\bibinfo{author}{\bibfnamefont{J.}~\bibnamefont{Tang}} \bibnamefont{and}
  \bibinfo{author}{\bibfnamefont{W.}~\bibnamefont{Winter}},
  \bibinfo{journal}{Phys. Rev.} \textbf{\bibinfo{volume}{D80}},
  \bibinfo{pages}{053001} (\bibinfo{year}{2009}),
  \eprint{arXiv:0903.3039}\relax
\mciteBstWouldAddEndPuncttrue
\mciteSetBstMidEndSepPunct{\mcitedefaultmidpunct}
{\mcitedefaultendpunct}{\mcitedefaultseppunct}\relax
\EndOfBibitem
\bibitem[{\citenamefont{Meloni et~al.}(2010)\citenamefont{Meloni, Ohlsson,
  Winter, and Zhang}}]{Meloni:2009cg}
\bibinfo{author}{\bibfnamefont{D.}~\bibnamefont{Meloni}},
  \bibinfo{author}{\bibfnamefont{T.}~\bibnamefont{Ohlsson}},
  \bibinfo{author}{\bibfnamefont{W.}~\bibnamefont{Winter}}, \bibnamefont{and}
  \bibinfo{author}{\bibfnamefont{H.}~\bibnamefont{Zhang}},
  \bibinfo{journal}{JHEP} \textbf{\bibinfo{volume}{04}}, \bibinfo{pages}{041}
  (\bibinfo{year}{2010}), \eprint{arXiv:0912.2735}\relax
\mciteBstWouldAddEndPuncttrue
\mciteSetBstMidEndSepPunct{\mcitedefaultmidpunct}
{\mcitedefaultendpunct}{\mcitedefaultseppunct}\relax
\EndOfBibitem
\bibitem[{\citenamefont{Fern{\'a}ndez-Mart{\'i}nez
  et~al.}(2007)\citenamefont{Fern{\'a}ndez-Mart{\'i}nez, Gavela,
  L{\'o}pez-Pav{\'o}n, and Yasuda}}]{FernandezMartinez:2007ms}
\bibinfo{author}{\bibfnamefont{E.}~\bibnamefont{Fern{\'a}ndez-Mart{\'i}nez}},
  \bibinfo{author}{\bibfnamefont{M.~B.} \bibnamefont{Gavela}},
  \bibinfo{author}{\bibfnamefont{J.}~\bibnamefont{L{\'o}pez-Pav{\'o}n}},
  \bibnamefont{and} \bibinfo{author}{\bibfnamefont{O.}~\bibnamefont{Yasuda}},
  \bibinfo{journal}{Phys. Lett.} \textbf{\bibinfo{volume}{B649}},
  \bibinfo{pages}{427} (\bibinfo{year}{2007}), \eprint{hep-ph/0703098}\relax
\mciteBstWouldAddEndPuncttrue
\mciteSetBstMidEndSepPunct{\mcitedefaultmidpunct}
{\mcitedefaultendpunct}{\mcitedefaultseppunct}\relax
\EndOfBibitem
\bibitem[{\citenamefont{Xing}(2008)}]{Xing:2007zj}
\bibinfo{author}{\bibfnamefont{Z.-z.} \bibnamefont{Xing}},
  \bibinfo{journal}{Phys. Lett.} \textbf{\bibinfo{volume}{B660}},
  \bibinfo{pages}{515} (\bibinfo{year}{2008}), \eprint{arXiv:0709.2220}\relax
\mciteBstWouldAddEndPuncttrue
\mciteSetBstMidEndSepPunct{\mcitedefaultmidpunct}
{\mcitedefaultendpunct}{\mcitedefaultseppunct}\relax
\EndOfBibitem
\bibitem[{\citenamefont{Goswami and Ota}(2008)}]{Goswami:2008mi}
\bibinfo{author}{\bibfnamefont{S.}~\bibnamefont{Goswami}} \bibnamefont{and}
  \bibinfo{author}{\bibfnamefont{T.}~\bibnamefont{Ota}},
  \bibinfo{journal}{Phys. Rev.} \textbf{\bibinfo{volume}{D78}},
  \bibinfo{pages}{033012} (\bibinfo{year}{2008}),
  \eprint{arXiv:0802.1434}\relax
\mciteBstWouldAddEndPuncttrue
\mciteSetBstMidEndSepPunct{\mcitedefaultmidpunct}
{\mcitedefaultendpunct}{\mcitedefaultseppunct}\relax
\EndOfBibitem
\bibitem[{\citenamefont{Xing and Zhou}(2008)}]{Xing:2008fg}
\bibinfo{author}{\bibfnamefont{Z.-z.} \bibnamefont{Xing}} \bibnamefont{and}
  \bibinfo{author}{\bibfnamefont{S.}~\bibnamefont{Zhou}},
  \bibinfo{journal}{Phys. Lett.} \textbf{\bibinfo{volume}{B666}},
  \bibinfo{pages}{166} (\bibinfo{year}{2008}), \eprint{arXiv:0804.3512}\relax
\mciteBstWouldAddEndPuncttrue
\mciteSetBstMidEndSepPunct{\mcitedefaultmidpunct}
{\mcitedefaultendpunct}{\mcitedefaultseppunct}\relax
\EndOfBibitem
\bibitem[{\citenamefont{Luo}(2008)}]{Luo:2008vp}
\bibinfo{author}{\bibfnamefont{S.}~\bibnamefont{Luo}}, \bibinfo{journal}{Phys.
  Rev.} \textbf{\bibinfo{volume}{D78}}, \bibinfo{pages}{016006}
  (\bibinfo{year}{2008}), \eprint{arXiv:0804.4897}\relax
\mciteBstWouldAddEndPuncttrue
\mciteSetBstMidEndSepPunct{\mcitedefaultmidpunct}
{\mcitedefaultendpunct}{\mcitedefaultseppunct}\relax
\EndOfBibitem
\bibitem[{\citenamefont{Altarelli and Meloni}(2009)}]{Altarelli:2008yr}
\bibinfo{author}{\bibfnamefont{G.}~\bibnamefont{Altarelli}} \bibnamefont{and}
  \bibinfo{author}{\bibfnamefont{D.}~\bibnamefont{Meloni}},
  \bibinfo{journal}{Nucl. Phys.} \textbf{\bibinfo{volume}{B809}},
  \bibinfo{pages}{158} (\bibinfo{year}{2009}), \eprint{arXiv:0809.1041}\relax
\mciteBstWouldAddEndPuncttrue
\mciteSetBstMidEndSepPunct{\mcitedefaultmidpunct}
{\mcitedefaultendpunct}{\mcitedefaultseppunct}\relax
\EndOfBibitem
\bibitem[{\citenamefont{Malinsk\'y et~al.}(2009)\citenamefont{Malinsk\'y,
  Ohlsson, and Zhang}}]{Malinsky:2009gw}
\bibinfo{author}{\bibfnamefont{M.}~\bibnamefont{Malinsk\'y}},
  \bibinfo{author}{\bibfnamefont{T.}~\bibnamefont{Ohlsson}}, \bibnamefont{and}
  \bibinfo{author}{\bibfnamefont{H.}~\bibnamefont{Zhang}},
  \bibinfo{journal}{Phys. Rev.} \textbf{\bibinfo{volume}{D79}},
  \bibinfo{pages}{073009} (\bibinfo{year}{2009}),
  \eprint{arXiv:0903.1961}\relax
\mciteBstWouldAddEndPuncttrue
\mciteSetBstMidEndSepPunct{\mcitedefaultmidpunct}
{\mcitedefaultendpunct}{\mcitedefaultseppunct}\relax
\EndOfBibitem
\bibitem[{\citenamefont{Blennow and
  Fern{\'a}ndez-Mart{\'i}nez}(2010)}]{Blennow:2009pk}
\bibinfo{author}{\bibfnamefont{M.}~\bibnamefont{Blennow}} \bibnamefont{and}
  \bibinfo{author}{\bibfnamefont{E.}~\bibnamefont{Fern{\'a}ndez-Mart{\'i}nez}},
  \bibinfo{journal}{Comput. Phys. Commun.} \textbf{\bibinfo{volume}{181}},
  \bibinfo{pages}{227} (\bibinfo{year}{2010}), \eprint{arXiv:0903.3985}\relax
\mciteBstWouldAddEndPuncttrue
\mciteSetBstMidEndSepPunct{\mcitedefaultmidpunct}
{\mcitedefaultendpunct}{\mcitedefaultseppunct}\relax
\EndOfBibitem
\bibitem[{\citenamefont{Antusch
  et~al.}(2009{\natexlab{b}})\citenamefont{Antusch, Blennow,
  Fern{\'a}ndez-Mart{\'i}nez, and L{\'o}pez-Pav{\'o}n}}]{Antusch:2009pm}
\bibinfo{author}{\bibfnamefont{S.}~\bibnamefont{Antusch}},
  \bibinfo{author}{\bibfnamefont{M.}~\bibnamefont{Blennow}},
  \bibinfo{author}{\bibfnamefont{E.}~\bibnamefont{Fern{\'a}ndez-Mart{\'i}nez}},
  \bibnamefont{and}
  \bibinfo{author}{\bibfnamefont{J.}~\bibnamefont{L{\'o}pez-Pav{\'o}n}},
  \bibinfo{journal}{Phys. Rev.} \textbf{\bibinfo{volume}{D80}},
  \bibinfo{pages}{033002} (\bibinfo{year}{2009}{\natexlab{b}}),
  \eprint{arXiv:0903.3986}\relax
\mciteBstWouldAddEndPuncttrue
\mciteSetBstMidEndSepPunct{\mcitedefaultmidpunct}
{\mcitedefaultendpunct}{\mcitedefaultseppunct}\relax
\EndOfBibitem
\bibitem[{\citenamefont{Rodejohann}(2009)}]{Rodejohann:2009cq}
\bibinfo{author}{\bibfnamefont{W.}~\bibnamefont{Rodejohann}},
  \bibinfo{journal}{Europhys. Lett.} \textbf{\bibinfo{volume}{88}},
  \bibinfo{pages}{51001} (\bibinfo{year}{2009}), \eprint{arXiv:0903.4590}\relax
\mciteBstWouldAddEndPuncttrue
\mciteSetBstMidEndSepPunct{\mcitedefaultmidpunct}
{\mcitedefaultendpunct}{\mcitedefaultseppunct}\relax
\EndOfBibitem
\bibitem[{\citenamefont{Malinsk{\'y} et~al.}(2009)\citenamefont{Malinsk{\'y},
  Ohlsson, Xing, and Zhang}}]{Malinsky:2009df}
\bibinfo{author}{\bibfnamefont{M.}~\bibnamefont{Malinsk{\'y}}},
  \bibinfo{author}{\bibfnamefont{T.}~\bibnamefont{Ohlsson}},
  \bibinfo{author}{\bibfnamefont{Z.-z.} \bibnamefont{Xing}}, \bibnamefont{and}
  \bibinfo{author}{\bibfnamefont{H.}~\bibnamefont{Zhang}},
  \bibinfo{journal}{Phys. Lett.} \textbf{\bibinfo{volume}{B679}},
  \bibinfo{pages}{242} (\bibinfo{year}{2009}), \eprint{arXiv:0905.2889}\relax
\mciteBstWouldAddEndPuncttrue
\mciteSetBstMidEndSepPunct{\mcitedefaultmidpunct}
{\mcitedefaultendpunct}{\mcitedefaultseppunct}\relax
\EndOfBibitem
\bibitem[{\citenamefont{He et~al.}(2009)\citenamefont{He, Oh, Tandean, and
  Wen}}]{He:2009ua}
\bibinfo{author}{\bibfnamefont{X.-G.} \bibnamefont{He}},
  \bibinfo{author}{\bibfnamefont{S.}~\bibnamefont{Oh}},
  \bibinfo{author}{\bibfnamefont{J.}~\bibnamefont{Tandean}}, \bibnamefont{and}
  \bibinfo{author}{\bibfnamefont{C.-C.} \bibnamefont{Wen}},
  \bibinfo{journal}{Phys. Rev.} \textbf{\bibinfo{volume}{D80}},
  \bibinfo{pages}{073012} (\bibinfo{year}{2009}),
  \eprint{arXiv:0907.1607}\relax
\mciteBstWouldAddEndPuncttrue
\mciteSetBstMidEndSepPunct{\mcitedefaultmidpunct}
{\mcitedefaultendpunct}{\mcitedefaultseppunct}\relax
\EndOfBibitem
\bibitem[{\citenamefont{Xing}(2009)}]{Xing:2009ce}
\bibinfo{author}{\bibfnamefont{Z.-z.} \bibnamefont{Xing}},
  \bibinfo{journal}{Phys. Lett.} \textbf{\bibinfo{volume}{B679}},
  \bibinfo{pages}{255} (\bibinfo{year}{2009}), \eprint{arXiv:0907.3014}\relax
\mciteBstWouldAddEndPuncttrue
\mciteSetBstMidEndSepPunct{\mcitedefaultmidpunct}
{\mcitedefaultendpunct}{\mcitedefaultseppunct}\relax
\EndOfBibitem
\bibitem[{\citenamefont{Chao et~al.}(2008)\citenamefont{Chao, Luo, Xing, and
  Zhou}}]{Chao:2007mz}
\bibinfo{author}{\bibfnamefont{W.}~\bibnamefont{Chao}},
  \bibinfo{author}{\bibfnamefont{S.}~\bibnamefont{Luo}},
  \bibinfo{author}{\bibfnamefont{Z.-z.} \bibnamefont{Xing}}, \bibnamefont{and}
  \bibinfo{author}{\bibfnamefont{S.}~\bibnamefont{Zhou}},
  \bibinfo{journal}{Phys. Rev.} \textbf{\bibinfo{volume}{D77}},
  \bibinfo{pages}{016001} (\bibinfo{year}{2008}),
  \eprint{arXiv:0709.1069}\relax
\mciteBstWouldAddEndPuncttrue
\mciteSetBstMidEndSepPunct{\mcitedefaultmidpunct}
{\mcitedefaultendpunct}{\mcitedefaultseppunct}\relax
\EndOfBibitem
\bibitem[{\citenamefont{Dev and Mohapatra}(2010)}]{Dev:2009aw}
\bibinfo{author}{\bibfnamefont{P.~S.~B.} \bibnamefont{Dev}} \bibnamefont{and}
  \bibinfo{author}{\bibfnamefont{R.~N.} \bibnamefont{Mohapatra}},
  \bibinfo{journal}{Phys. Rev.} \textbf{\bibinfo{volume}{D81}},
  \bibinfo{pages}{013001} (\bibinfo{year}{2010}),
  \eprint{arXiv:0910.3924}\relax
\mciteBstWouldAddEndPuncttrue
\mciteSetBstMidEndSepPunct{\mcitedefaultmidpunct}
{\mcitedefaultendpunct}{\mcitedefaultseppunct}\relax
\EndOfBibitem
\bibitem[{\citenamefont{Antusch et~al.}(2010)\citenamefont{Antusch, Blanchet,
  Blennow, and Fern{\'a}ndez-Mart{\'i}nez}}]{Antusch:2009gn}
\bibinfo{author}{\bibfnamefont{S.}~\bibnamefont{Antusch}},
  \bibinfo{author}{\bibfnamefont{S.}~\bibnamefont{Blanchet}},
  \bibinfo{author}{\bibfnamefont{M.}~\bibnamefont{Blennow}}, \bibnamefont{and}
  \bibinfo{author}{\bibfnamefont{E.}~\bibnamefont{Fern{\'a}ndez-Mart{\'i}nez}},
  \bibinfo{journal}{JHEP} \textbf{\bibinfo{volume}{01}}, \bibinfo{pages}{017}
  (\bibinfo{year}{2010}), \eprint{arXiv:0910.5957}\relax
\mciteBstWouldAddEndPuncttrue
\mciteSetBstMidEndSepPunct{\mcitedefaultmidpunct}
{\mcitedefaultendpunct}{\mcitedefaultseppunct}\relax
\EndOfBibitem
\bibitem[{\citenamefont{Rodejohann}(2010)}]{Rodejohann:2009ve}
\bibinfo{author}{\bibfnamefont{W.}~\bibnamefont{Rodejohann}},
  \bibinfo{journal}{Phys. Lett.} \textbf{\bibinfo{volume}{B684}},
  \bibinfo{pages}{40} (\bibinfo{year}{2010}), \eprint{arXiv:0912.3388}\relax
\mciteBstWouldAddEndPuncttrue
\mciteSetBstMidEndSepPunct{\mcitedefaultmidpunct}
{\mcitedefaultendpunct}{\mcitedefaultseppunct}\relax
\EndOfBibitem
\bibitem[{\citenamefont{Broncano
  et~al.}(2003{\natexlab{a}})\citenamefont{Broncano, Gavela, and
  Jenkins}}]{Broncano:2002rw}
\bibinfo{author}{\bibfnamefont{A.}~\bibnamefont{Broncano}},
  \bibinfo{author}{\bibfnamefont{M.~B.} \bibnamefont{Gavela}},
  \bibnamefont{and} \bibinfo{author}{\bibfnamefont{E.~E.}
  \bibnamefont{Jenkins}}, \bibinfo{journal}{Phys. Lett.}
  \textbf{\bibinfo{volume}{B552}}, \bibinfo{pages}{177}
  (\bibinfo{year}{2003}{\natexlab{a}}), \eprint{hep-ph/0210271}\relax
\mciteBstWouldAddEndPuncttrue
\mciteSetBstMidEndSepPunct{\mcitedefaultmidpunct}
{\mcitedefaultendpunct}{\mcitedefaultseppunct}\relax
\EndOfBibitem
\bibitem[{\citenamefont{Broncano
  et~al.}(2003{\natexlab{b}})\citenamefont{Broncano, Gavela, and
  Jenkins}}]{Broncano:2003fq}
\bibinfo{author}{\bibfnamefont{A.}~\bibnamefont{Broncano}},
  \bibinfo{author}{\bibfnamefont{M.~B.} \bibnamefont{Gavela}},
  \bibnamefont{and} \bibinfo{author}{\bibfnamefont{E.~E.}
  \bibnamefont{Jenkins}}, \bibinfo{journal}{Nucl. Phys.}
  \textbf{\bibinfo{volume}{B672}}, \bibinfo{pages}{163}
  (\bibinfo{year}{2003}{\natexlab{b}}), \eprint{hep-ph/0307058}\relax
\mciteBstWouldAddEndPuncttrue
\mciteSetBstMidEndSepPunct{\mcitedefaultmidpunct}
{\mcitedefaultendpunct}{\mcitedefaultseppunct}\relax
\EndOfBibitem
\bibitem[{\citenamefont{Abada et~al.}(2007)\citenamefont{Abada, Biggio, Bonnet,
  Gavela, and Hambye}}]{Abada:2007ux}
\bibinfo{author}{\bibfnamefont{A.}~\bibnamefont{Abada}},
  \bibinfo{author}{\bibfnamefont{C.}~\bibnamefont{Biggio}},
  \bibinfo{author}{\bibfnamefont{F.}~\bibnamefont{Bonnet}},
  \bibinfo{author}{\bibfnamefont{M.~B.} \bibnamefont{Gavela}},
  \bibnamefont{and} \bibinfo{author}{\bibfnamefont{T.}~\bibnamefont{Hambye}},
  \bibinfo{journal}{JHEP} \textbf{\bibinfo{volume}{12}}, \bibinfo{pages}{061}
  (\bibinfo{year}{2007}), \eprint{arXiv:0707.4058}\relax
\mciteBstWouldAddEndPuncttrue
\mciteSetBstMidEndSepPunct{\mcitedefaultmidpunct}
{\mcitedefaultendpunct}{\mcitedefaultseppunct}\relax
\EndOfBibitem
\bibitem[{\citenamefont{Gavela et~al.}(2009)\citenamefont{Gavela, Hambye,
  Hern{\'a}ndez, and Hern{\'a}ndez}}]{Gavela:2009cd}
\bibinfo{author}{\bibfnamefont{M.~B.} \bibnamefont{Gavela}},
  \bibinfo{author}{\bibfnamefont{T.}~\bibnamefont{Hambye}},
  \bibinfo{author}{\bibfnamefont{D.}~\bibnamefont{Hern{\'a}ndez}},
  \bibnamefont{and}
  \bibinfo{author}{\bibfnamefont{P.}~\bibnamefont{Hern{\'a}ndez}},
  \bibinfo{journal}{JHEP} \textbf{\bibinfo{volume}{09}}, \bibinfo{pages}{038}
  (\bibinfo{year}{2009}), \eprint{arXiv:0906.1461}\relax
\mciteBstWouldAddEndPuncttrue
\mciteSetBstMidEndSepPunct{\mcitedefaultmidpunct}
{\mcitedefaultendpunct}{\mcitedefaultseppunct}\relax
\EndOfBibitem
\bibitem[{\citenamefont{Ohlsson and Zhang}(2009)}]{Ohlsson:2008gx}
\bibinfo{author}{\bibfnamefont{T.}~\bibnamefont{Ohlsson}} \bibnamefont{and}
  \bibinfo{author}{\bibfnamefont{H.}~\bibnamefont{Zhang}},
  \bibinfo{journal}{Phys. Lett.} \textbf{\bibinfo{volume}{B671}},
  \bibinfo{pages}{99} (\bibinfo{year}{2009}), \eprint{arXiv:0809.4835}\relax
\mciteBstWouldAddEndPuncttrue
\mciteSetBstMidEndSepPunct{\mcitedefaultmidpunct}
{\mcitedefaultendpunct}{\mcitedefaultseppunct}\relax
\EndOfBibitem
\bibitem[{\citenamefont{Meloni et~al.}(2009)\citenamefont{Meloni, Ohlsson, and
  Zhang}}]{Meloni:2009ia}
\bibinfo{author}{\bibfnamefont{D.}~\bibnamefont{Meloni}},
  \bibinfo{author}{\bibfnamefont{T.}~\bibnamefont{Ohlsson}}, \bibnamefont{and}
  \bibinfo{author}{\bibfnamefont{H.}~\bibnamefont{Zhang}},
  \bibinfo{journal}{JHEP} \textbf{\bibinfo{volume}{04}}, \bibinfo{pages}{033}
  (\bibinfo{year}{2009}), \eprint{arXiv:0901.1784}\relax
\mciteBstWouldAddEndPuncttrue
\mciteSetBstMidEndSepPunct{\mcitedefaultmidpunct}
{\mcitedefaultendpunct}{\mcitedefaultseppunct}\relax
\EndOfBibitem
\bibitem[{\citenamefont{Donini et~al.}(2002)\citenamefont{Donini, Meloni, and
  Migliozzi}}]{Donini:2002rm}
\bibinfo{author}{\bibfnamefont{A.}~\bibnamefont{Donini}},
  \bibinfo{author}{\bibfnamefont{D.}~\bibnamefont{Meloni}}, \bibnamefont{and}
  \bibinfo{author}{\bibfnamefont{P.}~\bibnamefont{Migliozzi}},
  \bibinfo{journal}{Nucl. Phys.} \textbf{\bibinfo{volume}{B646}},
  \bibinfo{pages}{321} (\bibinfo{year}{2002}), \eprint{hep-ph/0206034}\relax
\mciteBstWouldAddEndPuncttrue
\mciteSetBstMidEndSepPunct{\mcitedefaultmidpunct}
{\mcitedefaultendpunct}{\mcitedefaultseppunct}\relax
\EndOfBibitem
\bibitem[{\citenamefont{Donini et~al.}(2009)\citenamefont{Donini, Fuki,
  L{\'o}pez-Pav{\'o}n, Meloni, and Yasuda}}]{Donini:2008wz}
\bibinfo{author}{\bibfnamefont{A.}~\bibnamefont{Donini}},
  \bibinfo{author}{\bibfnamefont{K.-i.} \bibnamefont{Fuki}},
  \bibinfo{author}{\bibfnamefont{J.}~\bibnamefont{L{\'o}pez-Pav{\'o}n}},
  \bibinfo{author}{\bibfnamefont{D.}~\bibnamefont{Meloni}}, \bibnamefont{and}
  \bibinfo{author}{\bibfnamefont{O.}~\bibnamefont{Yasuda}},
  \bibinfo{journal}{JHEP} \textbf{\bibinfo{volume}{08}}, \bibinfo{pages}{041}
  (\bibinfo{year}{2009}), \eprint{arXiv:0812.3703}\relax
\mciteBstWouldAddEndPuncttrue
\mciteSetBstMidEndSepPunct{\mcitedefaultmidpunct}
{\mcitedefaultendpunct}{\mcitedefaultseppunct}\relax
\EndOfBibitem
\bibitem[{\citenamefont{Mohapatra and Valle}(1986)}]{Mohapatra:1986bd}
\bibinfo{author}{\bibfnamefont{R.~N.} \bibnamefont{Mohapatra}}
  \bibnamefont{and} \bibinfo{author}{\bibfnamefont{J.~W.~F.}
  \bibnamefont{Valle}}, \bibinfo{journal}{Phys. Rev.}
  \textbf{\bibinfo{volume}{D34}}, \bibinfo{pages}{1642}
  (\bibinfo{year}{1986})\relax
\mciteBstWouldAddEndPuncttrue
\mciteSetBstMidEndSepPunct{\mcitedefaultmidpunct}
{\mcitedefaultendpunct}{\mcitedefaultseppunct}\relax
\EndOfBibitem
\bibitem[{\citenamefont{Bhattacharya et~al.}(2009)\citenamefont{Bhattacharya,
  Dey, and Mukhopadhyaya}}]{Bhattacharya:2009nu}
\bibinfo{author}{\bibfnamefont{S.}~\bibnamefont{Bhattacharya}},
  \bibinfo{author}{\bibfnamefont{P.}~\bibnamefont{Dey}}, \bibnamefont{and}
  \bibinfo{author}{\bibfnamefont{B.}~\bibnamefont{Mukhopadhyaya}},
  \bibinfo{journal}{Phys. Rev.} \textbf{\bibinfo{volume}{D80}},
  \bibinfo{pages}{075013} (\bibinfo{year}{2009}),
  \eprint{arXiv:0907.0099}\relax
\mciteBstWouldAddEndPuncttrue
\mciteSetBstMidEndSepPunct{\mcitedefaultmidpunct}
{\mcitedefaultendpunct}{\mcitedefaultseppunct}\relax
\EndOfBibitem
\bibitem[{\citenamefont{Blennow et~al.}(2010)\citenamefont{Blennow,
  Melb{\'e}us, Ohlsson, and Zhang}}]{Blennow:2010zu}
\bibinfo{author}{\bibfnamefont{M.}~\bibnamefont{Blennow}},
  \bibinfo{author}{\bibfnamefont{H.}~\bibnamefont{Melb{\'e}us}},
  \bibinfo{author}{\bibfnamefont{T.}~\bibnamefont{Ohlsson}}, \bibnamefont{and}
  \bibinfo{author}{\bibfnamefont{H.}~\bibnamefont{Zhang}}
  (\bibinfo{year}{2010}), \eprint{arXiv:1003.0669}\relax
\mciteBstWouldAddEndPuncttrue
\mciteSetBstMidEndSepPunct{\mcitedefaultmidpunct}
{\mcitedefaultendpunct}{\mcitedefaultseppunct}\relax
\EndOfBibitem
\end{mcitethebibliography}
\bibliographystyle{apsrevM}

\end{document}